\documentclass[12pt]{article}

\usepackage{epsfig}
\usepackage{graphicx}

\def\be{\begin{equation}}
\def\ee{\end{equation}}
\def\beq{\begin{eqnarray}}
\def\eeq{\end{eqnarray}}
\def\e{\epsilon}

\def\({\left (}
\def\){\right )}
\def\[{\left [}
\def\[{\right ]}

\begin{document}

\begin{titlepage}
\bigskip
\rightline{}
\rightline{gr-qc/0405050}
\bigskip\bigskip\bigskip\bigskip
\centerline{\Large \bf {Update on Cosmic Censorship Violation in AdS}}
\bigskip\bigskip
\bigskip\bigskip
 \centerline{\large Thomas Hertog${}^1$, Gary T. Horowitz${}^1$,
   and Kengo Maeda${}^2$}
        \bigskip\bigskip
 \centerline{\em ${}^1$ Department of Physics, UCSB, Santa Barbara, CA 93106}
  \centerline{\em hertog@vulcan.physics.ucsb.edu,
     gary@physics.ucsb.edu }
               \bigskip
  \centerline{\em ${}^2$ Department of General Education, Kobe City College
  of Technology}
  \centerline{\em 8-3 Gakuen-higashi-machi, Nishi-ku, Kobe 651-2194, Japan}
                \centerline{\em kmaeda@kobe-kosen.ac.jp}
	              \bigskip\bigskip

\begin{abstract}
We reexamine our proposed counterexample (gr-qc/0307102) to cosmic
censorship in anti de Sitter (AdS) space, and find a gap in the construction.
We mention some possible ways to close the gap, but at present
the question of whether cosmic censorship is violated in AdS remains open.
\end{abstract}

\end{titlepage}

\baselineskip=18pt

\setcounter{equation}{0}

We have recently pointed out 
that cosmic censorship  should
be easier to violate in asymptotically anti de Sitter (AdS) spacetimes 
\cite{Hertog03}.
The idea is simply the following. A
positive cosmological constant causes expansion, whereas a negative
cosmological constant causes contraction. So singularities are
easier to form in AdS. In particular, a homogeneous scalar field
rolling down a negative potential results in a curvature singularity.
At the same time, a large black hole (with Schwarzschild
radius $R_s$ larger than the AdS radius of curvature) requires a much larger
mass than the same size black hole in asymptotically flat spacetime. In four
dimensions, $M_{bh}\propto R_s^3$ rather than the familiar $M_{bh}\propto R_s$.
This
can be viewed as a result of the fact that the black hole must overcome
all the local negative energy density in its vicinity. Thus, compared with
asymptotically flat spacetimes,  singularities
are easier to form and black holes
are harder to form. This suggests
that it may be possible for finite mass localized initial data to
collapse to singularities, but not have enough mass to form a black hole
large enough to enclose the singularity.

In \cite{Hertog03} we claimed to find generic counterexamples 
to cosmic censorship along these lines. 
Some people have questioned our
result \cite{Alcubierre04,Garfinkle:2004sx}, but 
as we discuss below, these objections can be dealt with. However, we
have found a gap in our arguments which appears more serious. 
So, at the present time, 
the issue of cosmic censorship violation in localized gravitational
collapse in AdS remains open. 

We begin by briefly reviewing our proposed counterexamples.
We considered four dimensional
gravity coupled to a scalar field with potential $V(\phi)$
satisfying the positive energy theorem. The potential is bounded from 
below and has
a global minimum $-3 V_0 <0$ at
$\phi = 0$ and a local negative minimum $-3 V_1<0$
at $\phi = \phi_1 >0$.
We consider solutions that asymptotically approach the local (AdS) minimum
at $\phi_1$. Generally speaking, the positive energy theorem
holds if the barrier separating the extrema is
high enough, but it does not hold if the barrier is too low.

Consider a potential which is just on the verge of violating the positive
energy theorem (but still satisfies it). Then for any large radius $R$,
one can construct time symmetric and spherically symmetric
initial data with $\phi(0) =0$, $\phi(r>R) =\phi_1$ that have
total mass $M\propto R$ \cite{Hertog03}.
In other words, the configuration interpolates
between the global minimum and the local minimum and has a mass which only 
grows linearly with $R$. In general, the mass has
contributions proportional to $R^3$ as well as $R$. But by adjusting the
height of the potential 
barrier to be just on the verge of violating the positive
energy theorem,  and minimizing the contribution to the mass  
which depends on the potential, one finds that the $R^3$
contribution vanishes.

This minimal configuration depends only on the dimensionless ratio
$r/R$. It stays close to the global minimum for $r<\alpha R$ where $\alpha<1$
depends on $V$. However $\phi(r)>0$ for all $r>0$,
so it reaches
the global minimum only at the origin. This means that
 for large $R$, there is a large central region which is approximately
homogeneous with $\phi \ne 0$ almost everywhere. One expects that under
evolution, this central region will collapse to a singularity. 
If this singularity is eventually 
enclosed inside an event horizon, the radius of
the black hole would have to be proportional to $R$, and hence the mass
would have to be proportional to $R^3$. This is much larger than the available
mass which is only proportional to $R$. This solution is clearly very special
(since we have assumed spherically symmetric and time symmetric initial data)
but due to the large discrepancy in the mass, one could perturb the initial
data and still get a contradiction with the assumption that a black hole
will enclose the singularity.

There are two objections to this argument which have been raised in the 
literature. One is a numerical calculation by Garfinkle \cite{Garfinkle:2004sx}, in which he
numerically evolves our initial data and finds that a small black hole
forms. However, as he points out, he can only do the evolution for modest 
values of $R$. It is easy to check that for his value of $R$,
the size of the black hole required to enclose the singularity
is smaller than the AdS scale. Small black holes
in AdS only require a mass $M\propto R$ so there is no contradiction
between his findings and the claims we made.

Another objection has been raised by Alcubierre et al \cite{Alcubierre04}.
They
suggest that the scalar ``wall'' separating the regions where the scalar
field is near the global minimum and local minimum could continue to expand
indefinitely. They argue that a large Schwarzschild AdS black hole
could form in the central region because it is surrounded by a region of
space with negative energy density relative to infinity. They also do a
numerical evolution (with asymptotically flat boundary conditions)
which seems to support this possibility.
The problem with this argument is that Alcubierre et al \cite{Alcubierre04}
do not impose the
positive energy theorem. It is easy to show that in our case, this could
not occur. More precisely, the region where $\phi$ is close
to the global minimum cannot expand without increasing the total energy.
This is because our initial profile $\phi(r)$
is already chosen to minimize the potential (volume) contribution to the 
energy. Any
other shape for the ``wall'' will have higher energy. One can move the wall
out by increasing $R$, but since the energy
is proportional to $R$ this will increase the energy. Under evolution we 
expect our wall to spread out, moving in to smaller radii as well as 
expanding out. 

Nevertheless, there is a problem with our proposed counterexample.
It is not obvious that the central region will collapse to a singularity. 
The point is that even though homogeneous solutions collapse to a singularity 
for all initial $\phi=\phi_0>0$, the size of the initial homogeneous region 
one needs to ensure a singular evolution grows as $\phi_0 \rightarrow 0$. 
This is because one approaches perfect AdS in this limit, where light rays
can travel in from infinity in finite time.
In the central approximately homogeneous region, $r<\alpha R$, 
 the field $\phi$ sits everywhere close to the global 
minimum of $V$. Hence the spatial metric in this region will be approximately
\beq\label{sads}
ds^2 &=& {dr^2 \over  1+ V_0 r^2}  + r^2 d\Omega
\nonumber
\eeq
The proper distance from $r=0$ to $r=\alpha R$ is proportional to $\ln R$
for large $R$, but for any $\e>0$ the proper distance from $r=\e R$ to 
$r=\alpha R$ is only proportional to $\ln \alpha/\e$, independent of $R$ 
(for large $R$). Therefore, in order to make the approximately
homogeneous region 
$\e R<r<\alpha R$ larger, one must make $\e$ smaller. But this brings 
$\phi$ closer to the global minimum of the potential in the inner part of the 
central region, since $\phi$ is only a function of $r/R$.
So one cannot have an arbitrarily large region where $\phi$ is bounded away 
from the global minimum. The net result is that the homogeneous approximation
is probably not justified all the way to the singularity. 
(An early indication of this was seen for modest size bubbles 
in the numerical work described in \cite{Garfinkle:2004sx}. Here we are 
saying that even for arbitrarily large bubbles, there is no reason to 
expect that the interior will
evolve like a homogeneous solution.)

Another important recent development is due to Dafermos \cite{Dafermos:2004ws}.
Motivated by our work, 
he considered spherically symmetric solutions to gravity coupled to 
a scalar field with a potential $V(\phi)$ which is bounded from below. 
Dafermos showed that if
a solution did collapse to a spacelike singularity that could not be
enclosed inside an event horizon, the singularity cannot end or become
timelike. Instead, it would have to extend all the way to infinity and form
a Big Crunch. While this is a common occurrence in cosmology, it is a highly
unusual outcome for localized, finite energy initial data
in a theory satisfying the positive energy theorem. 
Nevertheless, the causal structure of AdS suggests that it might
be easier to form a Big Crunch here than in asymptotically flat spacetimes,
since signals can propagate to infinity in finite time.
One might call this possibility an AdS Crunch\footnote{Whether or not one views
an AdS Crunch as a violation of cosmic censorship depends on the definition
of cosmic censorship. The singularity is not naked, but nevertheless, one
cannot evolve for all time in the asymptotic region.}.
If it occurs, it can be viewed as a nonlinear instability of the AdS solution
in such theories. Recall that the positive energy theorem only guarantees 
stability of the AdS vacuum, and  does not rule out an AdS Crunch forming from 
finite energy excitations.

It is natural to ask if one could modify our construction to obtain 
initial data which do produce an AdS Crunch. The simplest possibility is
to modify our initial minimal configuration near the origin
so that for some $\e>0$, $\phi(r)$ is constant for $r<\e R$ and never reaches 
the global minimum. Then the central region will indeed be strictly 
homogeneous, and for large enough $R$, a singularity must form during
the evolution. However, the modification increases the mass by a term 
proportional to $R^3$ which turns out to be larger than the mass needed to 
form a black hole which encloses the singularity. So one needs a more
clever modification of the path or the potential.
If one could construct an initial configuration where the field is bounded
away from the global minimum while keeping $M\propto R$, then the evolution
would have to produce an AdS Crunch.

We have also argued \cite{Hertog03b} cosmic censorship can be violated 
in $N=8$ supergravity, starting with initial data of non-compact support.
This proposal has been criticized in \cite{Gutperle04} and in 
\cite{Hubeny:2004cn}, but we will return to this case elsewhere
\cite{Hertog04}.

\vskip 1cm

\centerline{{\bf Acknowledgments}}
This work was supported in part by NSF grant PHY-0244764.

 \end{document}